\documentclass[twocolumn,prl,aps,10pt]{revtex4-1}
\usepackage{graphicx,epsf}
\begin{document}

\title{Behavior of quantum entropies in polaronic systems}

\author{C.A. Perroni, V. Marigliano Ramaglia, and V. Cataudella}

\affiliation{CNR-SPIN  and Dipartimento di Scienze Fisiche, \\
Universit\`{a} degli Studi di Napoli ``Federico II'',\\
Complesso Universitario Monte Sant'Angelo,\\
Via Cintia, I-80126 Napoli, Italy}

\begin {abstract}
Quantum entropies and state distances are analyzed
in polaronic systems with short range (Holstein model) and long
range (Fr$\ddot{o}$hlich model) electron-phonon coupling. These
quantities are extracted by a variational wave function which
describes very accurately polaron systems with arbitrary size in all
the relevant parameter regimes. With the use of quantum information tools, the
crossover region from weak to strong coupling regime can be characterized with high precision.
Then, the linear entropy is found to be very sensitive to the range of the
electron-phonon coupling and the adiabatic ratio. Finally,
the entanglement entropy is studied as a function of the system size
pointing out that it not bounded, but scales as the logarithm of the
size either for weak electron-phonon coupling or for short range interaction.
This behavior is ascribed to the peculiar coupling induced by the single electron
itinerant dynamics on the phonon subsystem.

\end {abstract}

\maketitle

\newpage

\section {Introduction}

In the last years quantum entanglement has attracted enormous
interest as a key physical resource at the basis of quantum
information processing. \cite{horo} In particular, attention has
been devoted to clarify and quantify quantum entanglement in
many-body systems since entanglement measures provide insights on the
quantum correlations of many-body functions. \cite{fazio} For example,
entanglement entropy and its scaling properties are currently used
in order to better characterize quantum phase transitions.
In the case of local couplings between degrees of freedom,
the entropy of the reduced state of a subregion grows like the boundary area of
the subregion, and not like its volume, that is known as area
law. \cite{eisert} However, close to a transition, this law is violated
since the entropy becomes divergent as a function of the system size.
The size scaling very often follows a logarithm law.

Tools given by quantum information have been especially important
for mesoscopic systems where several schemes have been proposed
for detection and measurement of entanglement. \cite{been} Very recently,
a microwave-frequency mechanical oscillator has been cooled to its ground state with
high probability and coupled to a quantum bit. \cite{nature1,nature2} This coupling preserves the quantum states
and allows a time-domain control of the system. In these experiments the maximum
number of phonons in the relevant mechanical mode is very low, so that one expects that
the coupling between mesoscopic resonator and quantum bit is not strong. Moreover, a scenario
has been proposed in order to detect entanglement of a mechanical resonator and a qubit
in a nanoelectromechanic setup. \cite{schmidt}

The realization of such devices where many quantum bits or multiple electronic states
are coupled to mesoscopic or macroscopic mechanical objects is far from being easy. Therefore, it is important
to make theoretical studies of such systems elucidating the role of the coupling between electrons
and oscillators and their entanglement properties. In particular, the dimension of the system
becomes a very important parameter for the analysis. Instead of considering artificial devices,
one can analyze compounds already existing in nature where a array
of microscopic oscillators is present in mesoscopic or macroscopic systems.
Therefore, the aim of this work is to study from a quantum information perspective a system
relevant in many areas of condensed and nanoscopic matter: the polaron,
i.e. a single electron (with many accessible states in a bulk crystal or  a quantum dot)
interacting with lattice phonons. \cite{ranning}
It has been proved that it does not show any self-trapping
phase transition with increasing electron-phonon ($el-ph$) coupling , but a crossover
between solutions with small extended (large polaron) and large localized
(small polaron) lattice deformations. \cite{lowen}
Up to now, only quantum entropies have been analyzed for polaronic systems with small size and
short range ($SR$) $el-ph$ interactions.  \cite{zhao,stoja} In addition to
quantum entropies, in this work, distance measures are studied in
polaronic models with arbitrary size taking the free electron as reference state.
We will use a variational approach that is very accurate in all the coupling and adiabaticity
regimes. \cite{17,ourlong}  By means of these tools, the precise position of the crossover
region between weak and strong coupling regime is identified.

Polaron studies extend the analysis of simpler spin-boson models. \cite{kopp}
The entanglement for polaron is very interesting also for other reasons: it can be considered
as a measure of decoherence of the electronic state due to the coupling with phonons
or dephasing of the phononic state induced by the interaction with the single
electron. From this point of view, studies of polaronic entanglement are relevant
in mesoscopic systems such as quantum dots in polar semiconductors
that have been proposed as systems for quantum processing.
\cite{ferreira} Moreover, in these systems, the $el-ph$ interaction is not local.
Therefore, in this paper, we have for the first time analyzed the effect of
long range ($LR$) $el-ph$ couplings on entanglement
amount. It is found that  quantum entropies
are strongly dependent on the adiabatic ratio and range of the interaction.
The final part of the paper will focus on the size scaling of the
entanglement entropy. Even if the system is not critical,
the entropy of polaronic systems is unbounded, and it scales as the logarithm of the size
either for weak $el-ph$ coupling or for local interactions.

\section{Models and Variational Approach}

The prototype model with $SR$ local coupling is the well known Holstein one,\cite{4} while
that with $LR$ interactions is the Fr$\ddot{o}$hlich one.
\cite{froh} In particular a discrete version of the Fr$\ddot{o}$hlic model
will be studied. \cite{alex} The Hamiltonian of the polaronic systems is
\begin{eqnarray}
H=-t\sum_{<i,j>} c^{\dagger}_{i}c_{j} +\omega_0\sum_{i}
a^{\dagger}_i a_i  + H_{el-ph}. \label{1r}
\end{eqnarray}
In Eq.(\ref{1r}) $c^{\dagger}_{i}$ ($c_i$) denotes the electron
creation (annihilation) operator at site $i$, whose position
vector is indicated by $\vec{R}_{i}$, and the symbol $<>$ denotes
nearest neighbors linked through the transfer integral $t$. The
operator $a^{\dagger}_i$ ($a_i$) represents the creation
(annihilation) operator for phonon on the site $i$, and $\omega_0$
is the frequency of the optical local phonon modes. Both the $SR$ and
$LR$  models can be described by the general $el-ph$ Hamiltonian $H_{el-ph}$

\begin{eqnarray}
H_{el-ph}=\alpha \omega_0 \sum_{i,j}
f(|\vec{R}_i-\vec{R}_j|) c^{\dagger}_{i}c_{i}\left(
a_j+a^{\dagger}_j\right), \label{1ra}
\end{eqnarray}
where $\alpha$ controls the strength of $el-ph$ coupling, and
$f(|\vec{R}_i-\vec{R}_j| )$ is the interacting force between an
electron on the site $i$ and an ion displacement on the site $j$.
The units are such that the lattice parameter $a=1$ and $\hbar=1$.

The Hamiltonian (\ref{1r}) reduces to the Holstein model for
\begin{equation}
f(|\vec{R}_i-\vec{R}_j| )=\delta_{\vec{R}_i,\vec{R}_j},
\label{force1}
\end{equation}
while in the $LR$ case \cite {alex} the interaction
force is given by
\begin{equation}
f(|\vec{R}_i-\vec{R}_j|)=  \left(|\vec{R}_i-\vec{R}_j|^{2} +1
\right)^{-\frac{3}{2}}. \label{force}
\end {equation}

Through the matrix element $M_{\vec{q}}$, defined as the lattice
Fourier transform of $\alpha \omega_0 f(|\vec{R}_i|)$, one defines
the polaronic shift $E_p=\sum_{\vec{q}}
M^{2}_{\vec{q}}/\omega_0$ and the coupling constant
$\lambda=E_p/zt$, with $z$ lattice coordination number, that
represents a natural measure of the strength of the $el-ph$
coupling for any range of the interaction. Another important
parameter of polaronic systems is the adiabatic ratio
$\gamma=\omega_0/t$.

We adopt a variational approach previously proposed
\cite{17,ourlong} for the study of systems with variable range
$el-ph$ interactions and arbitrary size. Not only ground state energies, but also
effective masses and spectral weights calculated with this approach have been compared
with the results of numerical approaches finding excellent agreement.
The trial wave functions are
translational invariant Bloch states obtained by taking a superposition
of localized states centered on different lattice sites
\begin{equation}
|\psi^{(i)}_{\vec{k}}>=\frac{1}{\sqrt{N}}\sum_{\vec{R}_n}e^{i\vec{k}\cdot
\vec{R}_n}|\psi^{(i)}_{\vec{k}}(\vec{R}_n)>,
 \label{12rn}
\end{equation}
where
\begin{equation}
|\psi^{(i)}_{\vec{k}}(\vec{R}_n)> = e^{S^{(i)} ( \vec{R}_n )}
\sum_m \phi^{(i)}_{\vec{k}}(\vec{R}_m) c^{\dagger}_{m+n}|0>,
 \label{13rn}
\end{equation}
with
\begin{equation}
S^{(i)} ( \vec{R}_n ) = \sum_{\vec{q}}\left[
h^{(i)}_{\vec{q}}(\vec{k})a_{\vec{q}} e^{i\vec{q}\cdot \vec{R}_n}
+h.c.\right].
 \label{13rna}
\end{equation}
In the last equations, $N$ is the number of lattice sites
(corresponding to the dimensionality of the electron Hilbert
space), the apex $i=w,s$ indicates the weak and strong coupling
polaron wave function, respectively, $|0>$ denotes the electron
and phonon vacuum state, $h^{(i)}_{\vec{q}}(\vec{k})$ are the phonon
distribution functions and $\phi^{(i)}_{\vec{k}}(\vec{R}_m)$ are
variational parameters defining the spatial broadening of the
electronic wave function. For each function, the variational minimization becomes
accurate extending the electron wave function up to a few neighbors.

The ground state properties are determined by
considering as trial state $|\psi_{\vec{k}}>$ a linear
superposition of the weak and strong coupling wave functions
\begin{equation}
|\psi_{\vec{k}}>=\frac{A_{\vec{k}}
|\overline{\psi}^{(w)}_{\vec{k}}>+ B_{\vec{k}}
|\overline{\psi}^{(s)}_{\vec{k}}>}
{\sqrt{A^2_{\vec{k}}+B^2_{\vec{k}}
+2A_{\vec{k}}B_{\vec{k}}S_{\vec{k}}}}, \label{31r}
\end{equation}
where $|\overline{\psi}^{(i)}_{\vec{k}}>$ is the normalized wave
function and
$S_{\vec{k}}=<\overline{\psi}^{(w)}_{\vec{k}}|\overline{\psi}^{(s)}_{\vec{k}}>$
is the overlap factor of the two wave functions. In Eq.(\ref{31r})
$A_{\vec{k}}$ and $B_{\vec{k}}$ are two additional variational
parameters which provide the relative weight of the weak and
strong coupling solutions for any particular value of $\vec{k}$.
In the rest of this work, we will study the one-dimensional ground
state corresponding to $\vec{k}=k=0$ in the physically relevant
adiabatic regime $\gamma<1$.

\section{Results}

The main quantity extracted from the wave function is the
phonon-traced electron density operator $\rho_{el}$
\begin{equation}
\rho_{el}=Tr_{ph}\left[ |\psi_{k=0}> < \psi_{k=0}| \right],
\label{331r}
\end{equation}
where $Tr_{ph}$ denotes the trace over the phonon degrees of
freedom. In order to analyze the entanglement between electron and
phonon, one can use the linear entropy $S_L$
\begin{equation}
S_L=1-Tr_{el}\left[ (\rho_{el})^2 \right],
\label{332r}
\end{equation}
where $Tr_{el}$ stands for the trace over the electronic degrees
of freedom. \cite{nielsen} In Fig. 1, we report
the linear entropy as a function of the coupling $\lambda$ for
$SR$ and $LR$ interactions. It is zero for a free
electron, then it increases with $\lambda$ reaching
the saturation value $1-1/N$ that marks the transition to
the totally mixed state. For the Holstein model,
$S_L$ reaches the saturation value at a value of $\lambda$ slightly
larger than unity. In the $LR$ case, $S_L$ increases due to the
larger entanglement between electron and phonons. Moreover, the
crossover between the weak and strong coupling regimes is
smoother. \cite{ourlong} All these features make the linear
entropy a very important quantity to measure the change of polaron
features as function of the coupling $\lambda$.

\begin{figure}[htb]
\includegraphics[width=9cm,height=7cm]{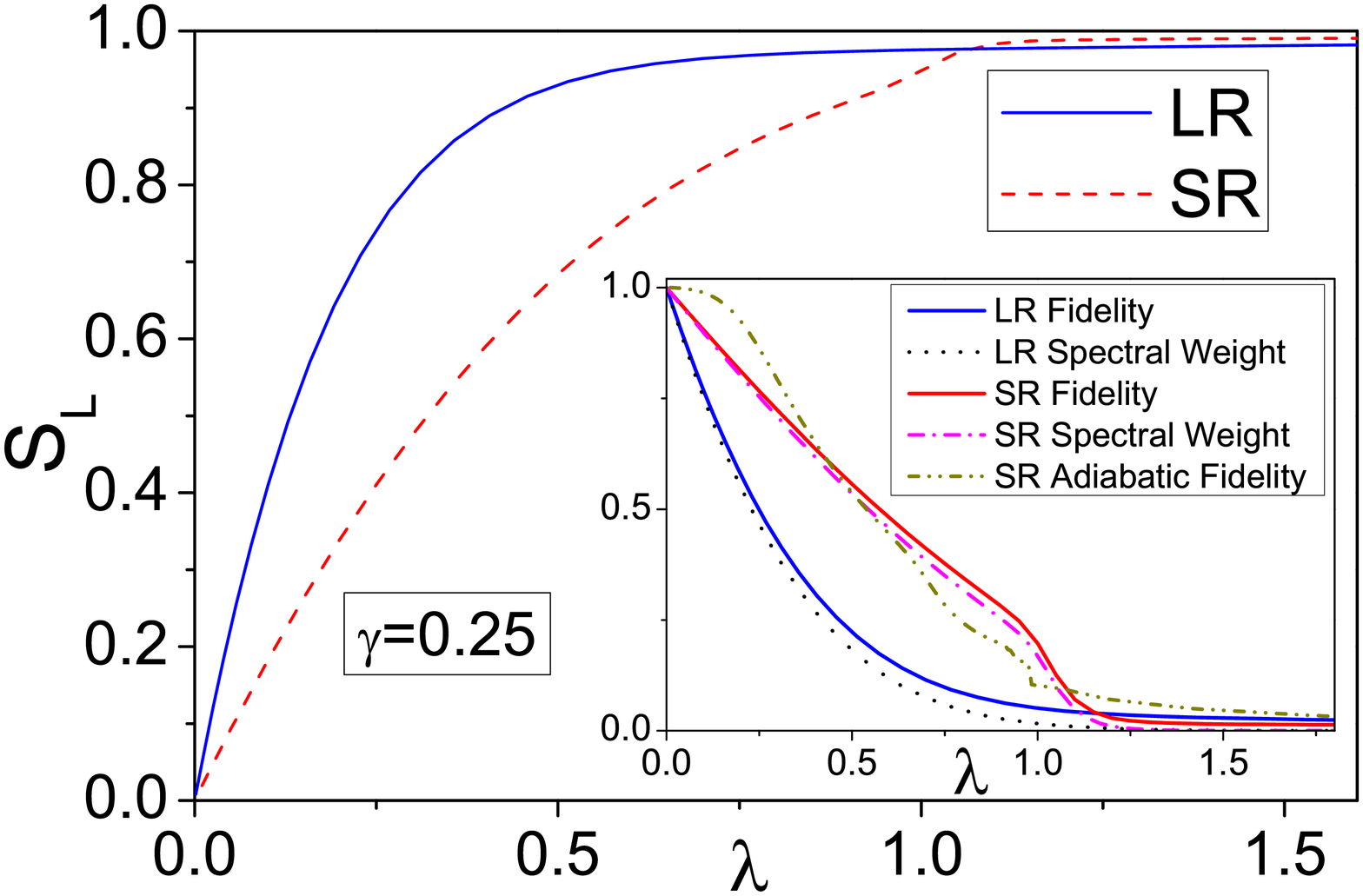}
\caption{Linear entropy as a function of $\lambda$ for $SR$
and $LR$ ranges of interaction at $\gamma=0.25$ and $N=128$. In
the inset, fidelity and spectral weight as a function of
$\lambda$. The fidelity in the fully adiabatic regime is also
shown.} \label{corrva}
\end{figure}

Another important quantity for the analysis of polaronic systems is the fidelity $F$
\begin{equation}
F=Tr_{el}\left[ (\rho_{free-el}) (\rho_{el}) \right], \label{333r}
\end{equation}
where $\rho_{free-el}$ is the free electron density operator
appropriate for periodic boundary conditions. \cite{schumacher} The
fidelity considered here is a measure of distance between the polaron and
the free electron state. In the inset of Fig.1, we
show the fidelity for $LR$ and $SR$ case. It is $1$ for
$\lambda=0$ and decreases with increasing $\lambda$ (for $SR$ it
is linear in the intermediate regime) up to a value close to zero
in the strong coupling regime. Therefore, in the maximally
entangled state, single electron and phonons are so strongly
coupled that free electron features have completely disappeared.
In the $LR$ case, due to the larger mixing between degrees of
freedom, the fidelity shows a marked tendency towards strong
coupling features.

In the inset of Fig.1, the fidelity is compared to the ground state spectral
weight $Z=|< \psi_{k=0}|c^{\dagger}_k=0|0>|^2$),
with $c^{\dagger}_k$ electron creation operator in the momentum
representation. $Z$ measures how much the quasi-particle is
different from the free electron ($Z = 1$). A small value of
it indicates a strong mixing of electronic and phononic degrees of
freedom. $Z$ and $F$ (in the same inset) share the same behavior as
function of $\lambda$ for both $SR$ and $LR$ case.

\begin{figure}[htb]
\includegraphics[width=9cm,height=7cm]{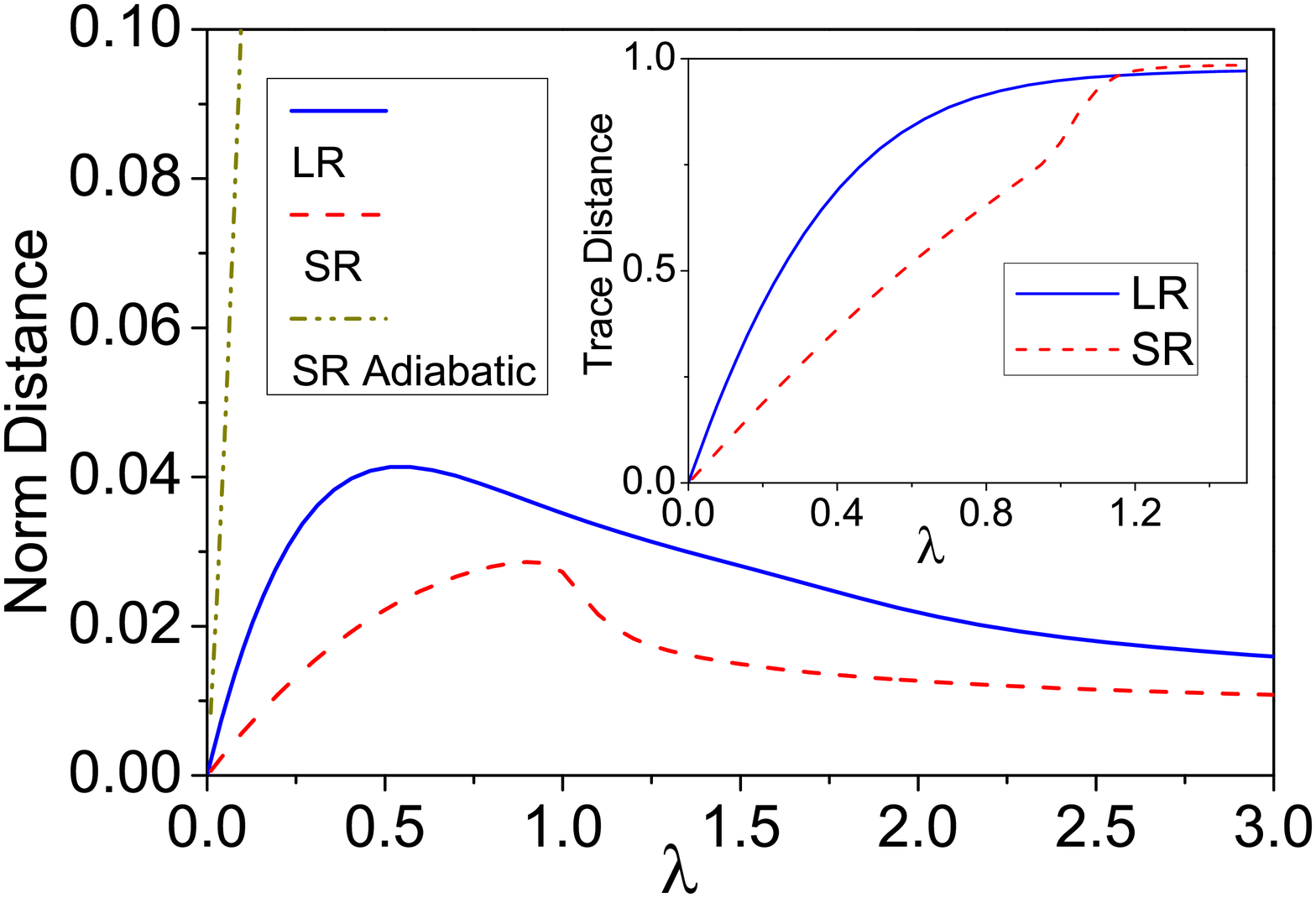}
\caption{Norm distance as function of $\lambda$ for $SR$ and
$LR$ $el-ph$ coupling at $\gamma=0.25$ and $N=128$. The same
quantity in the fully adiabatic regime is also shown. In the
inset, the trace distance as function $\lambda$ for the same
values of parameters.} \label{corrva1}
\end{figure}

The spectral weight has been used to distinguish qualitatively a
crossover regime ($0.1<Z< 0.9$) between the quasi-free-electron
one ($0.9<Z< 1$) and strong coupling one ($Z<0.1$).
\cite{17,ourlong} Our analysis shows that a new quantum measure, the norm distance,
is able to characterize in a quantitative way the crossover region. The norm distance is defined as the
eigenvalue norm of the difference density operator
$\sigma=\rho_{el}-\rho_{el-free}$: $||\sigma||_{\eta}=\max_i
|\eta_i|$, with $\eta_i$ eigenvalues of $\sigma$. \cite{nielsen} In Fig. 2,
the norm distance is reported for $SR$ and $LR$ couplings. In the first case, it is peaked at a value of $\lambda$
slightly smaller than unity. The $LR$ case it is much more
interesting since the distance shows a maximum at about
$\lambda=0.5$, which corresponds to an intermediate value of the
fidelity and spectral weight. Finally, it is possible to evaluate
the trace distance as the trace norm of $\sigma/2$. \cite{nielsen} This function
shows a behavior very similar to the linear entropy. From the
comparison of the two distances, it emerges that the maximum
eigenvector of $\sigma$ is always a fraction of the sum of all the
others. Therefore, the peak shown in the norm distance is to be
ascribed to the peculiar structure of the wave-function in this regime.

It is important to access the difference of the results between
adiabatic regime and fully adiabatic limit. In the limit
$\omega_0 \rightarrow 0$, the phonon fields are classical and the
only relevant $el-ph$ coupling is $\lambda$. Clearly, quantum
entropies vanish. However, it is still possible to study the
fidelity $F$ (shown in the inset of Fig.1). $F$ in this regime
shows an abrupt change at $\lambda=1$ for the $SR$ case. This
corresponds to the self-trapping transition towards a very
localized state which breaks the translation invariance. In Fig. 2,
we report the norm distance in the full adiabatic
limit for the $SR$ case. It is strongly larger than its
corresponding quantity with quantum phonons. Actually, in
strong coupling, one eigenvalue is close to unity, that
relative to the electron localized on a single site.

Since quantum entropies are zero in the fully adiabatic regime, we
analyze the behavior of the linear entropy close
to this limit. In Fig. 3, we report the linear entropy for the
$SR$ and $LR$ case as a function of the adiabatic ratio $\gamma$
for two values of the $el-ph$ constant $\alpha$. The entropies get
larger with increasing the adiabatic ratio. They are strongly
dependent on $\alpha$ and on the range of the $el-ph$ interaction.
Indeed, for large values of $\alpha$, the parameter $\lambda$
increases very fast as a function of $\gamma$. Therefore, the
linear entropy reaches the saturation value close to unity. Actually
quantum entropies are very sensitive to quantum phonon fluctuations,
so that they could be used as analyzer of the quantum nature of the oscillators.

\begin{figure}[htb]
\includegraphics[width=9cm,height=7cm]{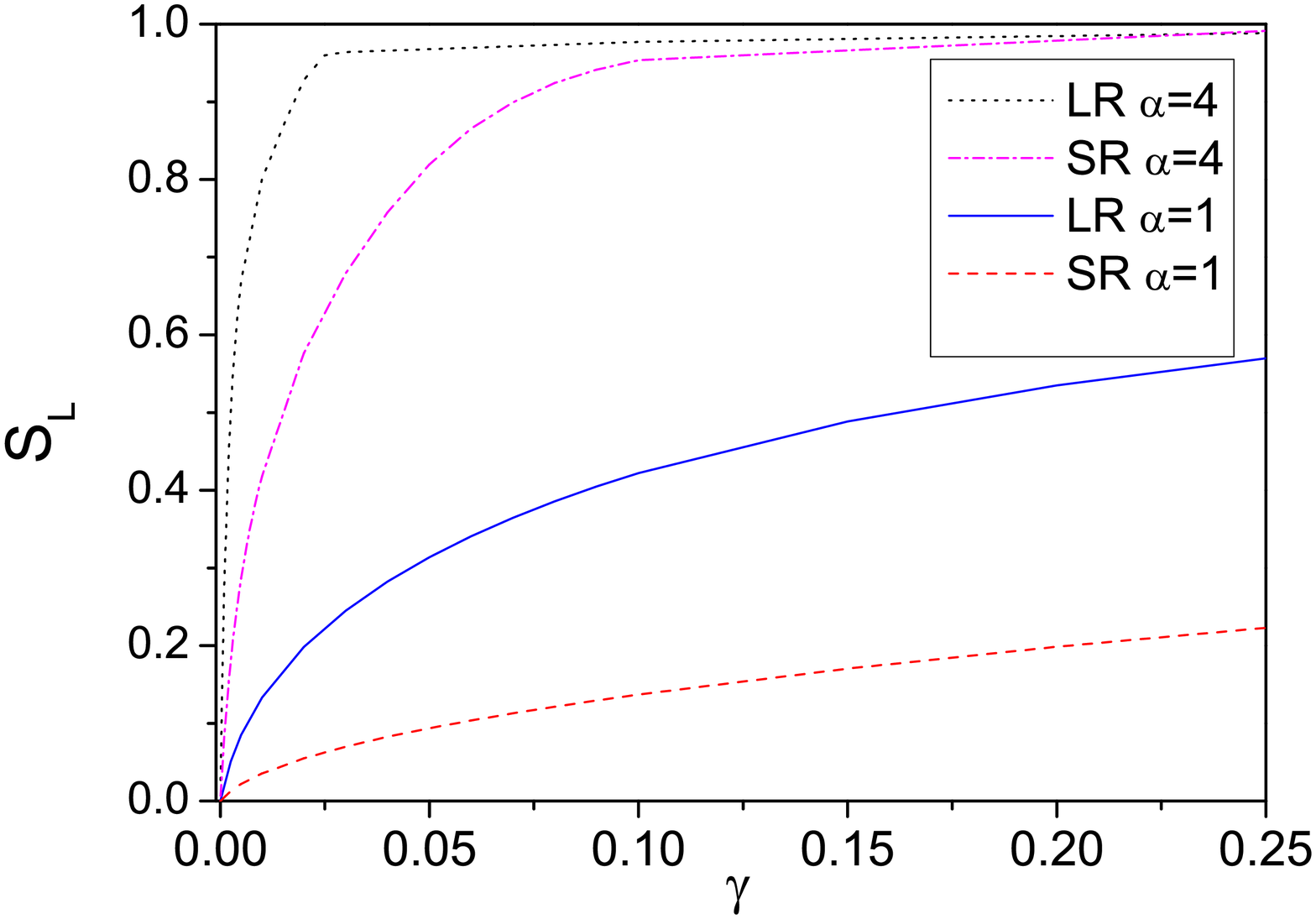}
\caption{Linear entropy as a function of the adiabatic ratio
$\gamma$ for several values of the $el-ph$ coupling constant
$\alpha$ at $N=128$.} \label{corrva1}
\end{figure}

In the last part of the paper, we will analyze the entanglement
or von Neumann entropy $S_{VN}$
\begin{equation}
S_{VN}=Tr_{el}\left[ (\rho_{el})ln (\rho_{el})
\right]=S_{VN}^{el}=S_{VN}^{phon},
\end{equation}
in particular its dependence on the size $N$. The equivalence between electron
and phonon entropy is due to the fact that the starting point is a
bipartite pure state. \cite{nielsen}
$S_{VN}$ is zero for decoupled electron and phonon degrees of
freedom, and has the maximum value of $ln(N)$, since $N$ is the
dimension of the smallest subsystem, the electron one.

At fixed size, the linear and von Neumann entropy are monotonic
functions of each other. Therefore, $S_{VN}$ increases as a
function of $\lambda$ up to a saturation value in the strong
coupling regime when the state is maximally entangled. We have found that,
in this limit, for both $SR$ and $LR$ interactions, $S_{VN}$ reaches
the maximum value, so that it is not bounded but scales as $ln(N)$.
This behavior in the strong coupling regime is due, in our opinion,
to the peculiar "long-range" coupling induced by the single electron itinerant dynamics on
the phonon subsystem.

\begin{figure}[htb]
\begin{center}
\includegraphics[width=9cm,height=7cm]{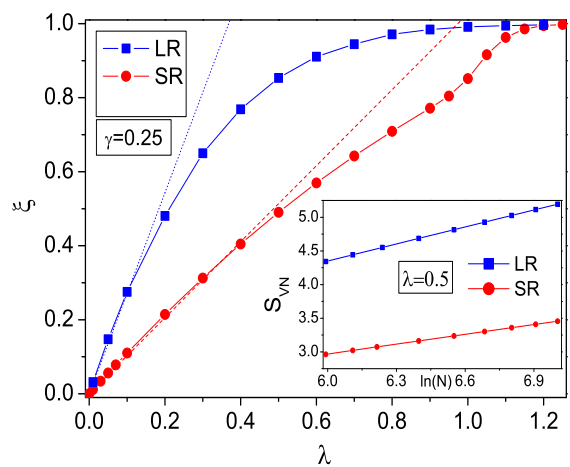}
\end{center}
\caption{The quantity $\xi$ as function of $\lambda$ at
$\gamma=0.25$. In the inset the von Neumann entropy as a function
of the system size $N$ at fixed value of $\lambda$.}
\label{corrva2}
\end{figure}

This result is even more surprising since it is not limited to the strong coupling limit.
The scaling proportional to $ln(N)$ is obtained also in the weak and intermediate regime.
At the second order of the perturbation theory in the $el-ph$ coupling, in the limit
of large $N$, $S_{VN} \rightarrow \bar{N}_{ph} ln(N)$,
where $\bar{N}_{ph}$ is the perturbative average number of excited virtual
phonons (linear as function of $\lambda$). We have found by means of an accurate fitting procedure
that, in all the regimes, $S_{VN}$ scales as $\xi ln(N)+a$, where $\xi$ (see Fig. 4) and $a$ depend on the
$el-ph$ coupling. In the inset of Fig. 4, we report the von Neumann entropies at fixed
value of $\lambda$ as function of the size $N$. With increasing the $el-ph$ coupling,
$\xi$ deviates from the linear dependence, and in the crossover regime it curves
towards the value of unity for strong coupling. Indeed, $\xi$ behaves as
the linear entropy shown in Fig.1.

One of the main results of this work is that the entanglement
entropy is unbounded and scales as a logarithm of the
size in all the regimes. This result is valid also for $LR$
$el-ph$ interaction. Actually, $S_{VN}$ cannot increase more than
$ln(N)$ due to the dimensional constraint of the electron Hilbert
space. This study allows to quantify the amount of the entanglement of
realistic systems, for example quantum dots in polar semiconductors,
where $LR$ polaronic effects can be important. In these systems, the
entanglement can be also related to the measure of decoherence of the
electronic state induced by the interactions with phonons. As a result of this study,
one can estimate that the entanglement entropy due to $el-ph$ coupling scales as the
logarithm of size with a proportionality constant that varies in a
simple way as a function of the parameters.

The experimental detection of entanglement is difficult for the bulk, but feasible
for nanostructures. Recently, an electronic measurement has been suggested in order to detect
entanglement between a qubit and an oscillator making use of an atomic point contact. \cite{schmidt}
Furthermore, again in the case of the interaction between qubit and resonator, time domain control
has been used in order to controllably create a phonon in the resonator and to observe
the exchange of this excitation between qubit and oscillator. \cite{nature2}
Effects of electric fields have been also analyzed for polarons in bulk semiconductors. \cite{gaal}
The pulse induces not only coherent lattice vibrations but also velocity drift oscillations of the electron. This
is again a demonstration of the more complex entanglement behavior between electron and phonon degrees of freedom
that has been the focus of this work.

We acknowledge R. Fazio for a critical reading of part of the manuscript.

\end{document}